\documentclass{ws-mpla}

\begin{document}

\markboth{J\'er\^ome Martin}
{Quintessence: a mini-review}

%%%%%%%%%%%%%%%%%%%%% Publisher's Area please ignore %%%%%%%%%%%%%%
\catchline{}{}{}{}{}
%%%%%%%%%%%%%%%%%%%%%%%%%%%%%%%%%%%%%%%%%%%%%%%%%%%%%%%%%%%%%%%%%%%

\title{QUINTESSENCE: A MINI-REVIEW}

\author{\footnotesize J\'ER\^OME MARTIN}

\address{Institut d'Astrophysique de Paris, UMR 7095-CNRS, 
Universit\'e Pierre et Marie Curie \\ 98bis boulevard Arago, 75014
Paris, France\\ jmartin@iap.fr}

\maketitle

\pub{Received (Day Month Year)}{Revised (Day Month Year)}

\begin{abstract}

Models where the accelerated expansion of our Universe is caused by a
quintessence scalar field are reviewed. In the framework of high energy
physics, the physical nature of this field is discussed and its
interaction with ordinary matter is studied and explicitly
calculated. It is shown that this coupling is generically too strong to
be compatible with local tests of gravity. A possible way out, the
chameleon effect, is also briefly investigated.

\end{abstract}

\ccode{PACS Nos.: 98.80.Cq, 98.70.Vc}

\section{Introduction}	
\label{sec:introduction}

It is now established that the expansion of our Universe is
accelerated~\cite{obs}. However, the reason for this acceleration is
still a theoretical mystery since there is no natural mechanism in the
standard model of cosmology, beside the cosmological constant, which
could explain this phenomenon. As is well-known, the cosmological
constant required in order to explain the observations is very far from
what one theoretically expects. This last fact reflects that the
acceleration is characterized by an energy scale of $\sim
10^{-3}\mbox{eV}$ which is very different from the natural scales of
particle physics where new physics could show up. Many proposal have
been made in order to explain this puzzle but it is fair to say that
none of them has won through. In this short review article, we consider
the case of quintessence, that is to say the case where a scalar field
is responsible for the accelerated expansion.

\par

This article is organized as follows. In
Sec.~\ref{sec:quintessenceinbrief}, we briefly review the main features
of quintessence while in Sec.~\ref{sec:modelbuilding} we discuss its
origin in high energy
physics. Sec.~\ref{sec:thequintessencefieldduringinflation} is devoted
to its behavior during inflation and
Sec.~\ref{sec:interactingquintessence} studies its coupling with
ordinary matter. Finally, in Sec.~\ref{sec:conclusions}, we present our
conclusions.

\section{Quintessence in brief}
\label{sec:quintessenceinbrief}

In this model, gravity is described by general relativity and the
Universe is made of radiation, pressure-less matter and quintessence, a
scalar field denoted $Q$ in what follows. Then, the Friedman equation
reads
\begin{eqnarray}
\frac{3}{a^2}{\cal H}^2 =\kappa \left[\frac{Q'^2}{2a^2}+V(Q)+\rho
_{_{\rm B}}\right]\, ,
\end{eqnarray}
where $\kappa \equiv 8\pi /m_{_{\rm Pl}}^2$, $m_{_{\rm Pl}}$ being the
Planck mass. The quantity $a(\eta)$ is the scale factor and ${\cal H}$
is defined by ${\cal H}=a'/a$, a prime denoting a derivative with
respect to conformal time. $\rho _{_{\rm B}}$ stands for the background
energy density and is either the radiation energy density or the
pressure-less energy density according to the era considered.  The
Friedman equation must be supplemented with two conservation equations
that read (here, we restrict ourselves to models of quintessence where
the kinetic term is standard~\cite{DBI})
\begin{eqnarray}
Q'' +2{\cal H}Q'+a^2\frac{{\rm d}V(Q)}{{\rm d}Q}=0\, ,
\quad 
\rho _{_{\rm B}}'+3{\cal H}\left(1+\omega _{_{\rm B}}\right)\rho
_{_{\rm B}} =0\, ,
\end{eqnarray}
where $\omega _{_{\rm B}}$ is the equation of state parameter defined to
be the pressure to energy density ratio. In the case of radiation
$\omega _{_{\rm B}}=1/3$ while $\omega _{_{\rm B}}=0$ for a pressure-less
fluid. Already at this stage, an important assumption is made, namely
that the quintessence scalar field and the background fluid are
separately conserved. We will come back to this crucial question in what
follows.

\par

The only quantity which has not yet been specified is the potential
$V(Q)$. Its shape should be chosen such that it gives a satisfactory
model. Obviously, the first requirement to be met is to have an
accelerated expansion, namely
\begin{equation}
\label{accela}
\frac{\ddot{a}}{a}=-\frac{\kappa }{6}
\left(\rho _{_{\rm cdm}}+2\rho _{_{\rm rad}}
+\rho _{_{\rm Q}}+3p_{_{\rm Q}}\right)>0\, ,
\end{equation}
where a dot means a derivative with respect to cosmic time. Clearly,
this can be achieved if the quintessence field presently dominates the
matter content of our Universe and if its potential $V(Q)$ is
sufficiently flat so that $p_{_{\rm Q}}<0$. However, this is not
sufficient. The acceleration should start relatively recently, {\it
i.e.} at a redshift of order one, $z_{_{\rm acc}}\sim {\cal O}(1)$,
which means that, prior to $z_{_{\rm acc}}$, $Q$ was subdominant or, in
other words, that $Q$ was a test field. Moreover, there is also the
question of the initial conditions. Which values $Q_{_{\rm ini}}$,
$Q_{_{\rm ini}}'$ should one take and at which initial time should one
impose them?

\par

A crucial observation is that the above mentioned questions can be
addressed if one assumes that the shape of $V(Q)$ is given by the
so-called Ratra-Peebles potential (typically, this is in fact more
general)~\cite{RP}
\begin{equation}
\label{eq:defpot}
V(Q)=M^{4+\alpha}Q^{-\alpha}\, ,
\end{equation}
where $M$ is a typical energy scale and $\alpha >0$ is a free positive
index. Indeed, under the assumption that the scalar field is a test
field, one can show that the Klein-Gordon (non-linear) equation
possesses a particular solution which is a scaling solution given by
\begin{equation}
\label{eq:attra}
Q\propto a^{3\left(1+\omega _{_{\rm B}}\right)/
\left(\alpha +2\right)}\, ,\quad 
\rho _{_{\rm Q}}\propto a^{-3\left(1+\omega _{_{\rm Q}}\right)}\, , 
\quad \omega _{_{\rm Q}}=
\frac{\alpha \omega _{_{\rm B}}-2}{\alpha +2}\,
.
\end{equation}
An interesting feature of this solution is that the equation of state
parameter $\omega _{_{\rm Q}}$ changes its value when one goes from the
radiation dominated epoch to the matter dominated epoch. In some sense,
the test scalar field adapts its behavior to the evolution of the scale
factor. One says that it tracks the background evolution and, for this
reason, we call it a tracker field. Another important property is that,
according to the above equations, $\rho _{_{\rm Q}}$ scales less rapidly
than $\rho _{_{\rm B}}$ and, hence, will eventually dominate the matter
content of the Universe (in which case, clearly, the test field
approximation breaks down).

\par

\begin{figure}[t]
\centerline{\psfig{file=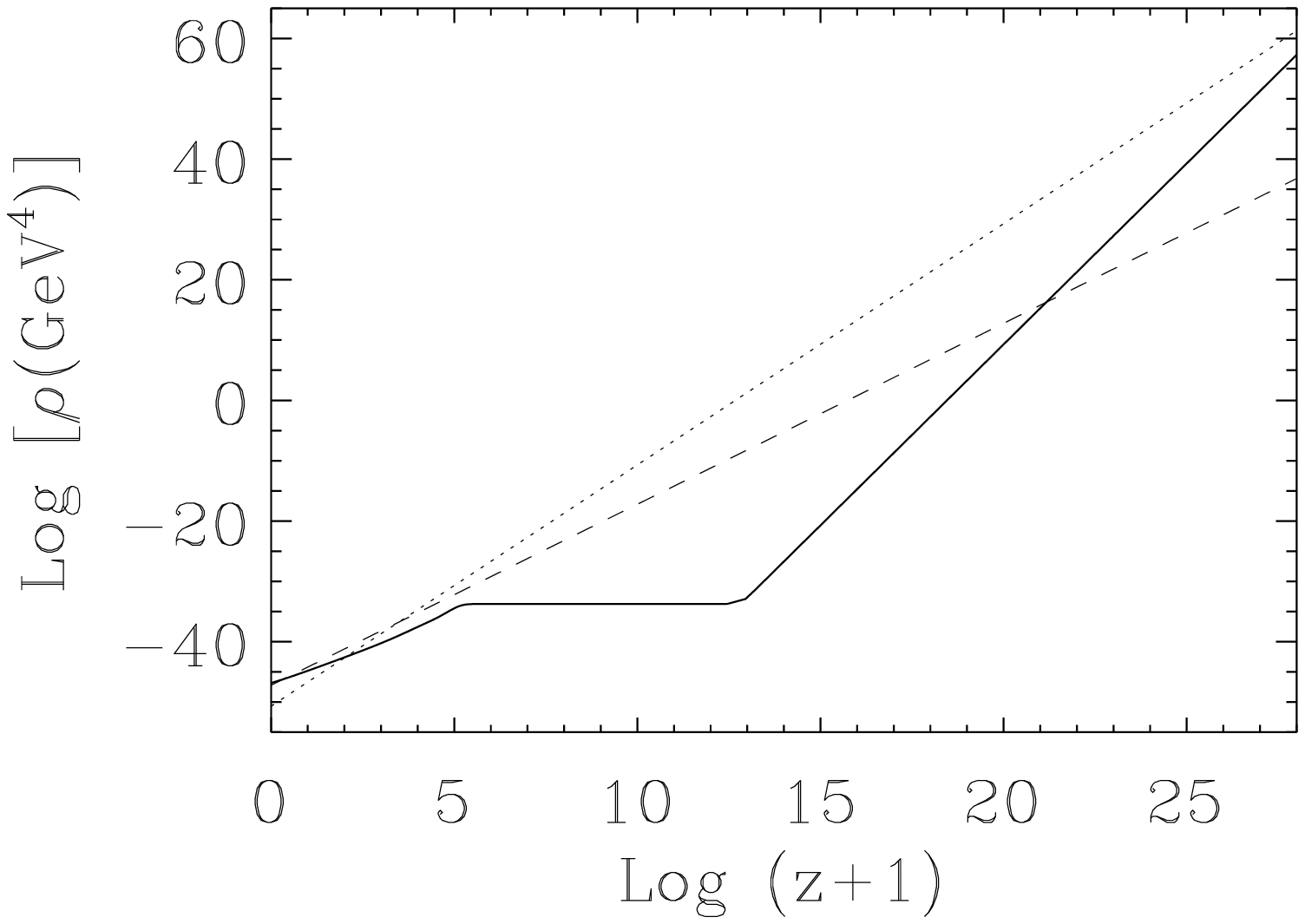,width=2.5in}
\psfig{file=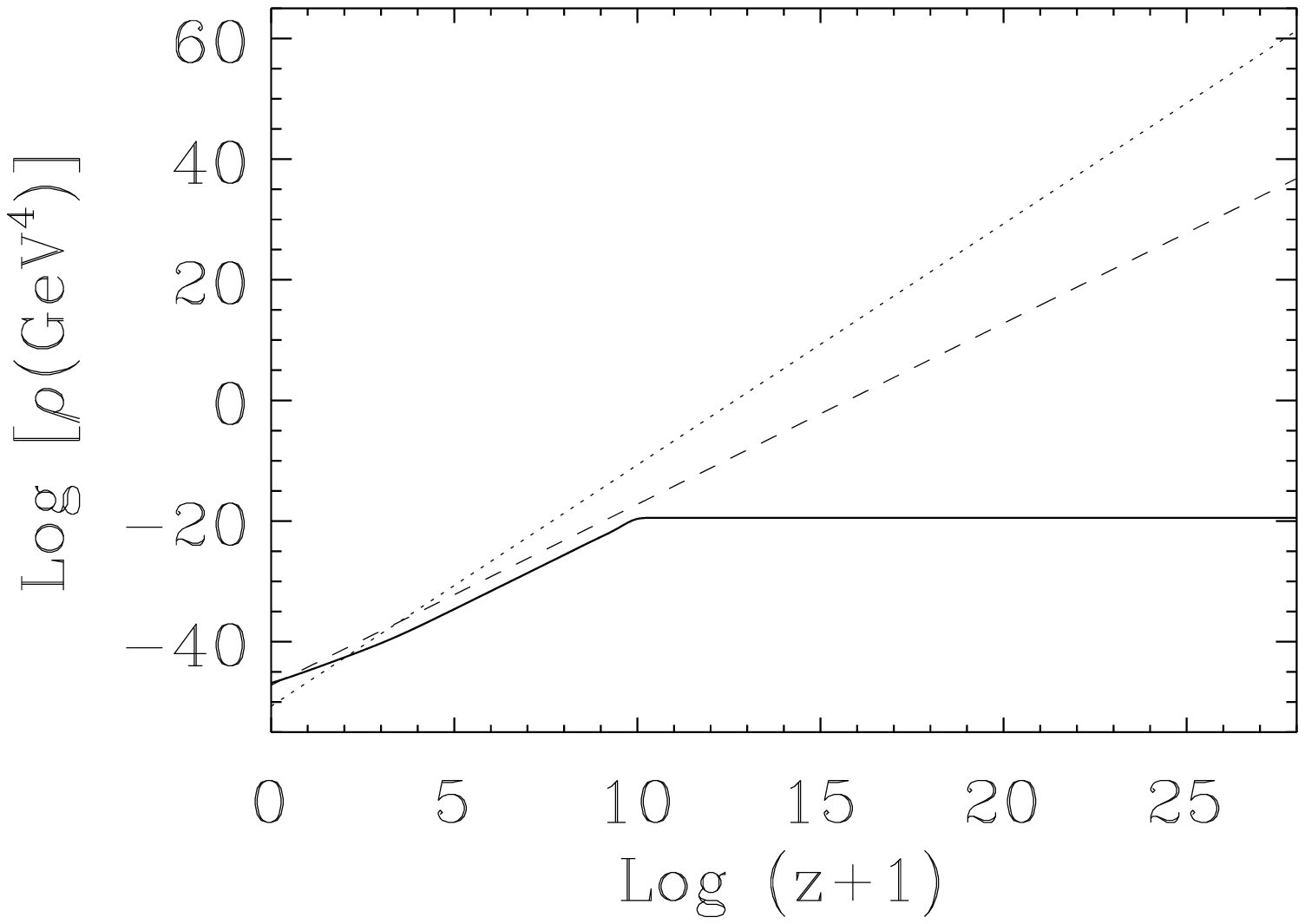,width=2.5in}}
\vspace*{8pt}
\caption{Evolution of radiation (dotted line), matter (dashed line) and 
quintessence (solid line) energy densities for two different 
initial values of $\rho _{_{\rm Q}}$ (left and right panels).
\label{fig:attra}}
\end{figure}

A priori, despite its nice features, this solution seems of little
importance because the chances to join it appears to involve a very
severe fine-tuning of the initial conditions. However, this is not so
because, beside being a scaling solution, it is also an attractor. In
practice, if we start the evolution, say, just after inflation, this
means that there is a huge range of initial conditions from which the
attractor is joined before present time. This property is illustrated in
Fig.~\ref{fig:attra} where the evolution of the radiation, matter and
quintessence energy densities is represented. The left and the right
panels corresponds to two different initial conditions for $\rho _{_{\rm
Q}}$ (chosen after reheating, at $z\sim 10^{28}$). On the left panel,
the attractor is joined at $z\sim 10^5$ while, on the right one, it is
joined at $z\sim 10^{10}$. Prior to the tracking regime, the evolution
of $\rho _{_{\rm Q}}$ depends on the initial conditions but, once the
field is on tracks, one always converges toward the same evolution. This
is why the solution~(\ref{eq:attra}), far from being useless, is on the
contrary very important. Indeed, almost regardless of the values of
$Q_{_{\rm ini}}$, $Q_{_{\rm ini}}'$, the evolution of $\rho_{_{\rm Q}}$
at small redshifts (but not too small) will always be described by
Eq.~(\ref{eq:attra}). Let us be more precise about this last
statement. In fact, one can show that the allowed initial values (the
initial values such that the attractor is joined before present time)
are approximatively such that $10^{-37}\mbox{GeV}^4<\rho _{_{\rm
Q}}<10^{61}\mbox{GeV}^4$ where $10^{-37}\mbox{GeV}^4$ corresponds to the
background energy density at equality whereas $10^{61}\mbox{GeV}^4$ is
the background energy density at reheating. This means that the initial
range encompasses about $100$ orders of magnitude in energy density!

We have already signaled that the solution~(\ref{eq:attra}) has the nice
feature to scale less quickly than the background and, hence, one is
guaranteed that, eventually, quintessence will dominate and cause the
acceleration of the expansion. Moreover, one can show that the redshift
at which the acceleration starts is given by
\begin{equation}
z_{_{\rm acc}}=
\left(\frac{\Omega _{_{\rm Q}}}{\Omega _{_{\rm
 m}}}\right)^{(2+\alpha)/6}-1\, ,
\end{equation}
and is always of order one given that $\Omega _{_{\rm Q}}\sim 0.7$ and
$\Omega _{_{\rm m}}\sim 0.3$. Therefore, one understands why the
acceleration started only recently.

\par

Let us also mention that cosmological perturbations can be computed in
this model. In particular, one can show that the quintessence field does
not cluster on scales smaller than the Hubble scale because the Jeans
mass is precisely given by $H^{-1}_0$. A complete study of the Cosmic
Microwave Background anisotropy has been carried out in
Ref.~\cite{BMR2}.

\par

\begin{figure}[t]
\centerline{\psfig{file=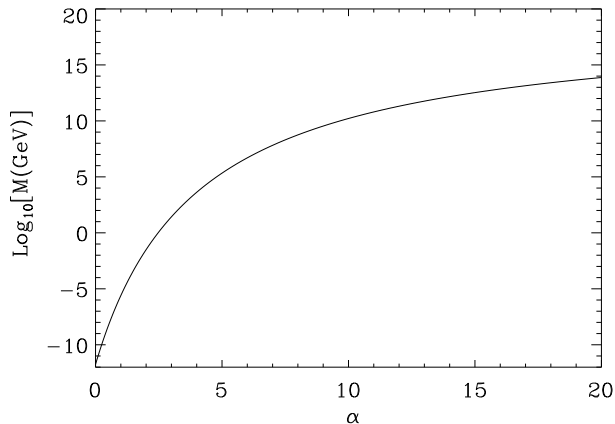,width=3.5in}}
\vspace*{8pt}
\caption{The quintessence energy scale $M$ as a function of the 
index $\alpha $ according to Eq.~(\ref{eq:scaleM}).\label{fig:scale}}
\end{figure}

It should be clear that all the nice above described properties are
obtained at the expense of requiring $\Omega _{_{\rm Q}}\sim 0.7$ which,
in turn, demands a (fine) tuning of the mass scale $M$ which is easy to
estimate. On the attractor the following relation holds
\begin{equation}
\frac{{\rm d}^2V}{{\rm d}Q^2}=\frac92 \frac{\alpha +1}{\alpha
}\left(1-\omega _{_{Q}}^2\right)H^2\, .
\end{equation}
The second derivative of the potential is nothing but the mass and can
roughly be written as $\sim V/Q^2$. On the other hand, when the field is
about to dominate, one has, using the Friedman equation, $H^2\sim
V/m_{_{\rm Pl}}^2$. Therefore, given that $\omega _{_{\rm Q}}$ and
$\alpha $ are of order one, the previous equation indicates that, at
present time, the vacuum expectation value of the quintessence field is
approximatively
\begin{equation} 
\langle Q \rangle _{_{\rm today}}\sim m_{_{\rm Pl}}\, .
\end{equation}
As a consequence, since the quintessence energy density is today of the
order of the critical energy density, one can write that $V\sim
M^{4+\alpha }m_{_{\rm Pl}}^{-\alpha}\sim \rho_{_{\rm cri}}$ and this
leads to
\begin{equation}
\label{eq:scaleM}
\log _{10}[M(\mbox{GeV})]\simeq \frac{19\alpha -47}{4\alpha +4}\, .
\end{equation}
One can check numerically that the above equation is a good
approximation. The corresponding plot is represented in
Fig.~\ref{fig:scale}. One notices that, for, say, $\alpha >6$, $M$ is
above the $\mbox{TeV}$ scale. What has been achieved is reminiscent of
the see-saw mechanism: thanks to the inverse power-law shape of the
Ratra-Peebles potential, one can justify the appearance of a very small
scale (by particle physics standards) even if the characteristic scale
of the problem is large. Therefore, it seems that we have gained
something.  Of course, the question is now to justify the inverse
power-law shape of the potential which was necessary in order to obtain
the nice properties described before. We turn to this question in the
next section.

\section{Model Building}
\label{sec:modelbuilding}

We have explained in the previous section how the
potential~(\ref{eq:defpot}) allows us to construct an interesting model
describing the accelerating Universe. It is clearly not sufficient to
postulate the existence of the quintessence field and to assume a shape
for its potential. One would like to understand the nature of this field
in high energy physics. Obviously, there is no candidate for the
quintessence field in the standard model of particle physics and,
therefore, one has to seek beyond, in the extensions of the standard
model. The most natural extensions are the super-symmetric ones and one
will focus on this class of models in the following.

\par

At this stage, the fact that $\langle Q \rangle _{_{\rm today}}\sim
m_{_{\rm Pl}}$ plays an important role~\cite{BM1,BM2,BMR1}. It means
that one should consider supergravity (SUGRA) models as opposed to
global super-symmetric theories~\cite{BM1,BM2}. Supergravity is a theory
where super-symmetry is gauged (made local) and its predictions usually
differ from those of global super-symmetry, the difference being of the
order of the vacuum expectation values of the fields in the problem
measured in Planck unit. Here, since $\langle Q \rangle _{_{\rm
today}}/m_{_{\rm Pl}}\sim 1$, these corrections are of order one and it
is mandatory to take them into account. This conclusion is very
important and has far-reaching consequences that will be discussed in
the rest of this article, the main issue at stake being whether the
assumptions exposed previously and necessary in order to build a
sensible quintessence model can be preserved in a SUGRA framework. We
will see that the answer to this question is negative.

\par

One can illustrate the previous claim on the following example. Can we
recover the potential~(\ref{eq:defpot}) in a SUGRA framework? We
postulate that the K\"ahler potential and the super-potential in the
quintessence sector can be Taylor expanded and are given
by~\cite{BM1,BM2}
\begin{eqnarray}
K\left(X,Y,Q\right) &=& XX^{\dagger }+QQ^{\dagger} +\kappa
^pYY^{\dagger} \left(QQ^{\dagger}\right)^p\, ,
\quad
W\left(X,Y,Q\right) =g X^2Y\, ,
\end{eqnarray}
Here $X$ and $Y$ are two charged fields under an (anomalous) $U(1)$
symmetry with charges $1$ and $-2$, while $Q$ is the neutral
quintessence field. The constant $g$ is a dimensionless coupling
constant and $p$ is a free coefficient. Then, one assumes that
\begin{equation}
\langle X \rangle= \xi \, ,\quad \langle Y \rangle =0\, .
\end{equation}
As a specific example, $\xi$ can be realized as a Fayet-Iloupoulous term
arising from the Green--Schwarz anomaly cancellation mechanism. In
supergravity, negative contributions to the scalar potential arise from
the vacuum expectation value of the super-potential. In the present
situation, $\left \langle W_{\rm quint} \right \rangle =0$ and we are
guaranteed that the potential is positive definite. We are now in a
position where the scalar potential can be computed. In supergravity, it
is given by $V={\rm e}^G \left(G^AG_A-3\right)/\kappa^2 $ and, in the
present situation, reduces to
\begin{equation}
V(Q)={\rm e}^{\kappa Q^2/2+\kappa \xi
^2}\frac{M^{4+2p}}{Q^{2p}}\, ,
\end{equation}
where the mass scale $M$ characterizing the potential can be expressed
as $M^{4+2p}\equiv 2^p g ^2\xi ^4 \kappa ^{-p}$. Firstly, we see that we
do not recover the potential~(\ref{eq:defpot}) but that the SUGRA
corrections play an important role: they have been exponentiated and
appear in the prefactor. Phenomenologically, it turns out to be an
advantage since the equation of state $\omega \equiv p_{_{\rm Q}}/\rho
_{_{\rm Q}}$ can be closer to $-1$ than with the Ratra--Peebles
potential. Secondly, this illustrates the fine-tuning of the parameters
in a new way. Indeed, we showed previously that the mass $M$ can be
large by particle physics standard which is certainly a nice
feature. However, we see here that there is still a somehow hidden
fine-tuning problem. Indeed, if one uses the expression of the mass
obtained before and assume that the coupling constant $g $ is of order
one (in order to avoid a fine-tuning), then one has $\xi \sim \rho
_{_{\rm cri}}^{1/4}$, {\it i.e.} a tiny scale. Therefore, despite the
satisfactory value of $M$, there is still a severe fine-tuning of
$\langle X\rangle$.

\section{The Quintessence field during inflation}
\label{sec:thequintessencefieldduringinflation}

We have seen that one of the main advantage of the quintessence scenario
is that it is insensitive to the choice of the initial conditions. More
precisely, if the energy density after inflation, at the beginning of
the subsequent radiation-dominated era, is such that
$10^{-37}\mbox{GeV}^4<\rho _{_{\rm Q}}<10^{61}\mbox{GeV}^4$, then the
attractor is joined. It is a huge range but, nevertheless, it is finite.
Since this range refers to the value of the quintessence energy density
after inflation, one can wonder how $Q$ evolves during inflation and
whether this evolution can drive $\rho _{_{\rm Q}}$ away from the
allowed range.

\par

As we are going to see, many effects can influence the behavior of $Q$
during inflation. For the purpose of illustration, let us assume that
the inflaton potential is given by $V_{\rm inf}=V_0\left(\phi /m_{_{\rm
Pl}}\right)^n$ (large field model) and that $Q$ and $\phi$ do not
interact. Almost by definition, $Q$ is a test field during inflation and
its evolution can be found by solving the Klein-Gordon equation. One
arrives at 
\begin{equation}
\label{qsolnneq2}
\frac{Q}{m_{_{\rm Pl}}} = 
\Biggl\{\left(\frac{Q_{_{\rm ini}}}{m_{_{\rm Pl}}}\right)^{\alpha +2}
+\frac{\alpha (\alpha +2)}{n(2-n)} \frac{M^{4+\alpha}
m_{_{\rm Pl}}^{-\alpha}}{V_0}\left[\left(
\frac{\phi_{_{\rm ini}}}{m_{_{\rm Pl}}}\right)^{2-n}
-\left(\frac{\phi }{m_{_{\rm Pl}}}\right)^{2-n}
\right]\Biggr\}^{1/(\alpha +2)}\, ,
\end{equation}
for $n\neq 2$ while, for $n=2$, the result reads
\begin{equation}
\label{qsoln2}
  \frac{Q}{m_{_{\rm Pl}}} = 
\left[\left(\frac{Q_{_{\rm ini}}}{m_{_{\rm Pl}}}\right)^{\alpha +2}
  +\frac{\alpha (\alpha +2)}{2}\frac{M^{4+\alpha}
m_{_{\rm Pl}}^{-\alpha}}{V_0}
\ln\frac{\phi_{_{\rm ini}}}{\phi}\right]^{1/(\alpha +2)}\, .
\end{equation}
As a matter of fact, for any value of $n$ we obtain that $Q\simeq
Q_{_{\rm ini}}$ at all times, that is to say the field is frozen, if the
initial value satisfies the constraint
\begin{equation}
\label{constQini}
\left(\frac{Q_{_{\rm ini}}}{m_{_{\rm Pl}}}\right)^{\alpha +2}
\gtrsim\left(\frac{M^{4+\alpha}m_{_{\rm Pl}}^{-\alpha}}{V_0}\right)
\left(\frac{\phi _{_{\rm ini}}}{m_{_{\rm Pl}}}\right)^{2-n}\, .
\end{equation}
In this case the initial conditions at the beginning of inflation
directly corresponds to those after reheating.

\par

Unfortunately, things are not that simple. There are at least two other
effects that are a priori present. The first effect is that the quantum
behavior of the quintessence field during inflation must be taken into
account~\cite{ML,stocha}. Indeed the various quantum kicks undergone by
the quintessence field during inflation could be such that, at the end
of inflation, $Q$ has drifted away too much and is outside the allowed
range or, more precisely, has a low probability of falling in this
allowed range. This can be checked by means of the stochastic inflation
formalism. In this approach the quintessence field becomes a stochastic
quantity the evolution of which is controlled by the Langevin equation
\begin{equation}
\label{eq:Langevin}
\frac{{\rm d}Q}{{\rm d}t}+\frac{V'(Q)}{3H(\phi)} =
\frac{H^{3/2}(\phi)}{2\pi}\xi_{_{\rm Q}}(t)\, ,
\end{equation}
where $\xi_{_{\rm Q}}$ is the quintessence white-noise field such that
$\langle\xi_{_{\rm Q}}(t)\xi_{_{\rm Q}}(t')\rangle=\delta(t-t')$. Of
course, the quantum effects for the inflaton field must also be taken
into account. This means that $\phi $ is also a stochastic quantity
described by its own Langevin equation. However, one can show that its
influence is not important~\cite{stocha}.

\begin{figure}[t]
\centerline{\psfig{file=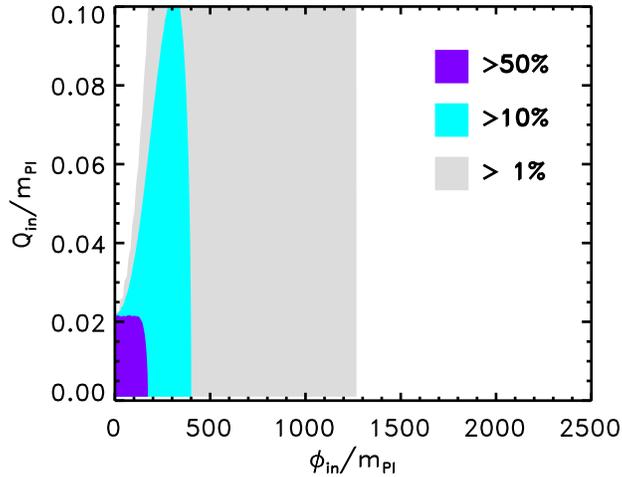,width=3.5in}}
\vspace*{8pt}
\caption{Probability contours in the initial conditions plane $(Q_{_{\rm
ini}},\phi_{_{\rm ini}})$ of having the quintessence vacuum expectation
value in the allowed range at the end of inflation. The plot has been
obtained for $n=2$ and $\alpha =6$. \label{fig:contour_alpha6}}
\end{figure}

In Ref.~\cite{stocha}, the Langevin equation for the quintessence field
has been solved for the Ratra-Peebles potential and the mean value,
variance and probability distribution function of $Q$ have been
computed. Then, a given model will be accepted if a large part of its
probability distribution calculated at the end of inflation is contained
within the allowed range, that is to say
\begin{equation}
\frac{Q_{\rm min}}{m_{_{\rm Pl}}}
\equiv 10^{-107/\alpha }<\frac{Q}{m_{_{\rm Pl}}}<\frac{Q_{\rm
 max}}{m_{_{\rm Pl}}}\equiv 10^{-9/\alpha }\, .
\end{equation}
The corresponding probability has been computed in Ref.~\cite{stocha}
and is presented in Fig.~\ref{fig:contour_alpha6} for $n=2$ and $\alpha
=6$. The main result is that there is an upper bound of the initial
value of the inflaton field if one wants the quintessence field to be on
tracks today. This is equivalent to a constraint on the total number of
e-folds during inflation, very roughly speaking
\begin{equation}
N_{_{\rm T}}\lesssim 10^{20(\alpha-2)/[\alpha(n+2)]}\, .
\end{equation}
This can be easily understood. If inflation lasts too long, then the
quantum quintessence field undergoes the quantum kicks during a very
long period and, therefore, the probability of falling within the
allowed range is small. The conclusion is that having $Q$ on tracks
today put relatively stringent constraints on the initial conditions at
the beginning of inflation. Hence, when inflation is taken into account,
the insensitivity to the initial conditions of the quintessence scenario
does not seem to be as efficient as usually claimed.

\par

There is still another effect which must be taken into account. So far,
we have assumed that the quintessence field and the inflaton are not
coupled. However, in SUGRA, as we shall see, this is not a reasonable
assumption and $Q$ and $\phi$ must necessarily
interact~\cite{quintinf}. Since we have argued that quintessence must be
described in SUGRA, for consistency, inflation must also be described in
this framework. In addition, we consider here large field models and the
fact that inflation occurs for $\phi>m_{_{\rm Pl}}$ reinforces the
previous argument. In the inflaton sector, we consider a class of models
described by the following K\"ahler potential
\begin{eqnarray}
\label{eq:kahlerinf}
K_{\rm inf} =-\frac{3}{\kappa }\ln \left[\kappa ^{1/2}\left(\rho+ \rho
^{\dagger }\right)-\kappa {\cal K}\left(\phi-\phi^\dagger
\right)\right]+{\cal G}\left(\phi-\phi^\dagger\right)\, ,
\end{eqnarray}
where ${\cal K}$ and ${\cal G}$ are given by
\begin{eqnarray}
{\cal K} &=& -\frac{1}{2}\left(\phi -\phi ^{\dagger }\right)^2\, , \quad
{\cal G}=+\frac{1}{2}\left(\phi -\phi ^{\dagger }\right)^2\, ,
\end{eqnarray}
and where $\rho $ represents, for instance, a modulus of a string
compactification. The super-potential $W_{\rm inf}=W_{\rm inf}\left(\rho
,\phi \right)$ is taken to be
\begin{eqnarray}
W_{\rm inf}(\rho, \phi ) &=& \frac{\alpha }{2}m\phi^2 \, .
\end{eqnarray}
Then, straightforward calculations lead to
\begin{equation}
V_{\rm inf}(\rho ,\phi)=\frac{1}{\Delta ^2(3-\Delta )}\alpha ^2m^2\phi
^2\, ,
\end{equation}
where $\Delta =\kappa^{1/2}(\rho +\rho ^\dagger )$. It is easy to see
that the modulus can be stabilized if $\Delta =2$. In this case, the
potential takes the form
\begin{equation}
V_{\rm inf}(\phi)=\frac{\alpha ^2}{4} m^2\phi^2 \, ,
\end{equation}
which is nothing but the usual chaotic inflation potential if one
chooses $\alpha =\sqrt{2}$. 

\par

Then, the most simple assumption is that the quintessence and inflation
sectors are decoupled, {\it i.e.}  that the total K\"ahler potential and
super-potential can be written as
\begin{eqnarray}
K &=& K_{\rm quint}\left(X,Y,Q\right)+K_{\rm inf}\left(\rho ,\phi
\right)\, , \quad W =  W_{\rm quint}\left(X,Y,Q\right)+W_{\rm
inf}\left(\rho ,\phi\right)\, ,
\end{eqnarray}
where the quintessential K\"ahler potential and super-potential have been
given in the previous section. The next step consists in computing the
scalar potential. Applying the standard SUGRA formalism one obtains $
V\left(\phi ,Q \right)=V_{\rm inf}+V_{\rm quint}+V_{\rm inter}$
with~\cite{quintinf}
\begin{equation}
V_{\rm inter}\propto \frac{m^2}{m_{_{\rm Pl}}^4}\phi ^4Q^2\, .
\end{equation}
As announced, we have a coupling between the inflaton and the
quintessence field. This simple calculation is interesting because it
exemplifies many interesting properties: even if the two fields live in
different sectors, they interact. Technically, this is because, in the
formula $V={\rm e}^G \left(G^AG_A-3\right)/\kappa^2 $, the $G$
multiplies $G^AG_A$. Physically, this is because, in SUGRA, the various
fields always interact through gravity even if they live in different
sectors. This is why the coupling constant is the Planck mass and why
the coupling is said to be Planck suppressed (see the $m_{_{\rm
Pl}}^{-4}$ in the above equation). However, because $\langle \phi
\rangle \sim m_{_{\rm Pl}}$, the coupling is not small.

\par

What is the effect of the coupling on the evolution of $Q$ during
inflation? Does it drive $Q$ away from the allowed range? From the
quintessence point of view, there is a time dependent effective
potential given by $V_{\rm quint}+V_{\rm inter}$. One can show that,
after a transitory regime, $Q$ quickly settles at the bottom of this
potential and, then, just follows the time-dependent minimum of the
effective potential
\begin{equation}
\label{eq:Qmin}
Q_{\rm min}(N)=m_{_{\rm Pl}}\left\{\frac{\alpha}{32 \pi
^2}\left(\frac{H_0}{m_{_{\rm Pl}}}\right)^2 
\left(\frac{m}{m_{_{\rm Pl}}}\right)^{-2}
\left[\frac{\phi (N)}{m_{_{\rm Pl}}}\right]^{-4} \right\}^{1/(\alpha+2)}\, .
\end{equation}
This is true regardless of the initial conditions at the beginning of
inflation. The above solution is thus an attractor and this sets the
initial conditions for the quintessence field after inflation. For
$\alpha=6$, which is our fiducial model, one has at the end of
inflation: $Q_{\rm min}(N=N_{_{\rm T}}) \simeq 1.9\times
10^{-14}m_{_{\rm Pl}}$. The effect of the non-renormalizable interaction
is therefore to force the quintessence field to remain small during
inflation. This time, the conclusion is positive since, typically, these
values are within the allowed range.

\section{Interacting Quintessence}
\label{sec:interactingquintessence}

We have just seen that, because SUGRA is ``universal'', the inflaton
field and the quintessence field necessarily interact. Clearly, this
conclusion calls into question our basic starting assumption, namely
that quintessence today does not interact with the rest of the world,
see Sec.~\ref{sec:quintessenceinbrief} where the quintessence
stress-energy tensor was separately conserved. We expect the same causes
to produce the same effects: the quintessence field must interact with
the ordinary matter fields~\cite{BMpart,BMcosmo}. The question is now to
prove it (rather than just assuming it in a toy model) and to compute
explicitly the form of this interaction for a given model. The most
generic approach is to consider that there are three different sectors
in the theory: the observable (where ordinary -- electrons, quarks,
Higgs etc ... -- matter and cold dark matter live), hidden (where
super-symmetry is broken) and quintessence sectors. As a consequence,
the hypothesis of separate sectors implies that the K\"ahler and super
potentials are given by the following expressions
\begin{eqnarray}
K &=& K_{\rm quint}+K_{\rm hid}+K_{\rm obs}\, , \quad W = W_{\rm
quint}+W_{\rm hid}+W_{\rm obs}\, .
\end{eqnarray}
Again, despite that quintessence and ordinary matter live in separate
sectors, one expects them to interact. In addition, one expects their
interaction to be controlled by $m_{_{\rm Pl}}$ and to be Planck
suppressed. However, since $\langle Q\rangle \sim m_{_{\rm Pl}}$ today,
this does not mean that the strength of this interaction will be
small. This interaction can be computed explicitly if we are given a
model, for instance, if one conservatively assumes the observable sector
to be described by the mSUGRA model~\cite{Nilles}. All the corresponding
consequences have been studied in detail in
Refs.~\cite{BMpart,BMcosmo}. Here, we just recap some of the main
conclusions.

\par

An important effect is that the electroweak symmetry breaking is
affected by the presence of the quintessence field. In the mSUGRA model,
there are two $\mbox{SU}(2)_{\rm L}$ Higgs doublets
\begin{equation}
H_{\rm u}=\begin{pmatrix} H_{\rm u}^+ \cr H_{\rm u}^0 \end{pmatrix} \,
, \quad H_{\rm d}=\begin{pmatrix} H_{\rm d}^0 \cr H_{\rm d}^-
\end{pmatrix}\, ,
\end{equation}
that have opposite hyper-charges, $Y_{\rm u}=1$ and $Y_{\rm d}=-1$ with a
super-potential given by $W_{\rm obs}=\mu H_{\rm u}\cdot H_{\rm
d}$. Usually, the fermions acquire a mass through their interaction with
the Higgs bosons. The mass is proportional to the Yukawa coupling and to
Higgs vacuum expectations values. In presence of quintessence, because
of the unavoidable interaction of $Q$ with $H_{\rm u}$ and $H_{\rm d}$,
the vacuum expectation values of $H_{\rm u}$ and $H_{\rm d}$, $v_{\rm
u}$ and $v_{\rm d}$, become $Q$-dependent (that is to say
time-dependent), namely
\begin{eqnarray}
\label{vuvd}
v_{\rm u}(Q)&=&\frac{v(Q)\tan \beta (Q)}{\sqrt{1+\tan^2 \beta (Q)}}\, 
\quad v_{\rm d}(Q) = \frac{v(Q)}{\sqrt{1+\tan ^2\beta (Q)}}\, ,
\end{eqnarray}
where $v\equiv \sqrt{v_{\rm u}^2+v_{\rm d}^2}\sim 174\mbox{GeV}$ and
$\tan \beta \equiv v_{\rm u}/v_{\rm d}$. It implies that fermion masses
become $Q$ dependent. Moreover there are two kinds of masses, depending
on whether the fermions couple to $H_{\rm u}$ or $H_{\rm d}$
\begin{eqnarray}
m_{{\rm u}}^{_{\rm F}}(Q)&=& \lambda_{{\rm u}}^{_{\rm F}} {\rm
  e}^{\kappa K_{\rm quint}/2+\sum _i\vert a_i\vert ^2/2}v_{\rm u}(Q)\,
  , \\  m_{{\rm d}}^{_{\rm F}}(Q)&=&\lambda_{{\rm d}}^{_{\rm F}}
  {\rm e}^{\kappa K_{\rm quint}/2+\sum _i\vert a_i\vert ^2/2}v_{\rm
  d}(Q)\, ,
\end{eqnarray}
where $\lambda_{{\rm u},a}^{_{\rm F}}$ and $\lambda_{{\rm d},a}^{_{\rm
F}}$ are the Yukawa coupling of the particle coupling either to $H_{\rm
u}$ or $H_{\rm d}$. The coefficients $a_i$ describe the hidden sector,
see Ref.~\cite{BMpart,BMcosmo} for more details. The time-dependence of
the fermions mass has drastic consequences that we briefly review in the
following.

\par

It was claimed before that, once a model is given, the form of the
interaction between quintessence and the rest of the world can be
determined. Let us illustrate this on the example where the quintessence
sector is described by the model presented in
Sec.~\ref{sec:quintessenceinbrief}. Then, one can
demonstrate~\cite{BMcosmo} that $\tan \beta (Q)$ can be expressed as
\begin{equation}
\label{tanbetasugra}
\tan \beta(Q) \simeq \frac{\delta _1+\delta _2\kappa Q^2+ \delta_3
  \kappa^2 Q^4}{\delta
_4+\delta_5 \kappa Q^2}\left[1+\sqrt{1+\frac{\left(\delta _4+\delta
_5\kappa Q^2\right)^2}{\left(\delta _1+\delta _2\kappa
Q^2+ \delta_3 \kappa^2 Q^4\right)^2}}\right]\, ,
\end{equation}
where the coefficients $\delta _1$, $\delta _2$, $\delta _3$, $\delta_4$
and $\delta _5$ can easily be evaluated in terms of the physical
parameters characterizing the model from the previous equations, $\mu$,
$m_{3/2}^0$ (gravitino mass) and $m_{1/2}^0$ (scalar mass) given at the
GUT scale. The expression of the scale $v(Q)$ can also be obtained from
the minimization of the Higgs potential along the lines described in
Ref.~\cite{BMpart}. One obtains
\begin{equation}
\label{v}
v(Q)=\frac{2}{\sqrt{g^2+g'{}^2}}{\rm e}^{\kappa K_{\rm quint}/2}
\sqrt{\left\vert \left \vert \mu \right \vert ^2+m_{H_{\rm
u}}^2\right\vert} +{\cal O}\left(\frac{1}{\tan \beta }\right)\, ,
\end{equation}
with the following expression for $m_{H_{\rm u}}$, coming from the
renormalization group equations, see Ref.~\cite{BS}
\begin{eqnarray}
\label{mhu}
m^2_{H_{\rm u}}\left( Q\right) &=& m^2_{H_{\rm d}}(Q)
-0.36\left(1+\frac{1}{\tan^2 \beta}\right)
\Biggl\{\left(m_{3/2}^0\right)^2\left(1-
\frac{1}{2\pi}\right)+8\left(m_{1/2}^0\right)^2 \nonumber \\ & & +
\left(0.28 -\frac{0.72}{\tan ^2\beta }\right )\left[A(Q) + 2
  m^0_{1/2}\right]^2\Biggr\}\, ,\\
\label{mhd}
m^2_{H_{\rm d}}\left(Q\right) &=&
      \left(m_{3/2}^0\right)^2\left(1-\frac{0.15}{4\pi}\right) +
      \frac{1}{2} \left(m^0_{1/2}\right)^2\, ,
\end{eqnarray}
and 
\begin{equation}
A(Q)=M_{_{\rm S}}\left(1+\frac{\kappa Q^2}{3}\right)\, ,\quad
B(Q)=M_{_{\rm S}}\left(1+\frac{\kappa Q^2}{2}\right)\, ,
\end{equation}
$M_{_{\rm S}}$ being the super-symmetric breaking scale. Notice that $A$
and $B$ follow the same universal relationship as in the mSUGRA model
despite the presence of the quintessence field. As promised, everything
is now explicit. One may notice that the coupling appears to be much
more complicated than the simple forms usually assumed in the
literature.

\par

Let us now investigate the consequences of having time-dependent fermion
masses.  Firstly, there will be a fifth force. If the mass of the
quintessence field is less than $10^{-3}\mbox{eV}$, the range of the
force is such that it can be experimentally seen. In order not to to be
in contradiction with fifth force experiments such as the recent Cassini
spacecraft experiment, its strength must be small and one must require
the parameter $\alpha _{\rm u,d}$ defined by
\begin{eqnarray}
\label{alpha}
\label{eq:defalpha}
\alpha_{\rm u,d}(Q) &\equiv & \left\vert \frac{1}{\kappa^{1/2}}
\frac{{\rm d}\ln m_{\rm u,d}^{_{\rm F}}(Q)}{{\rm d} Q}\right \vert
\end{eqnarray}
to be such that $\alpha_{\rm u,d}^2\le 10^{-5}$~\cite{GR}. This result
is valid for a gedanken experiment involving the gravitational effects
on elementary particles. For macroscopic bodies, the effects are more
subtle. Generically, in the models presented above, this limit is
violated and quintessence is in trouble. 

\par

Secondly, we have violations of the weak equivalence principle. This is
due to the fact that, in the mSUGRA model, the fermions couple
differently to the two Higgs doublets $H_{\rm u}$ and $H_{\rm
d}$. Violations of the weak equivalence principle are quantified in
terms of the $\eta_{_{\rm AB}}$ parameter defined by~\cite{GR}
\begin{equation}
\eta_{_{\rm AB}} \equiv \left(\frac{\Delta a}{a}\right)_{_{\rm AB}}
=2\frac{a_{_{\rm A}}-a_{_{\rm B}}}{a_{_{\rm A}}+a_{_{\rm B}}}\, ,
\end{equation}
for two test bodies A and B in the gravitational background of a third
one E.  Current limits~\cite{GR} indicate that $\eta_{_{\rm AB}}
=(+0.1\pm 2.7\pm 1.7)\times 10^{-13}$. Again, this limit applies only
for $m_{_{\rm Q}}<10^{-3}\mbox{eV}$.

\par

Thirdly, another consequence of the interaction between dark energy and
the observable sector is the variation of the gauge couplings, depending
on the complexity of the underlying model, more details on this specific
question can be found in Ref.~\cite{BMpart}.

\par 

Fourthly, another consequence of having $Q$-dependent masses is that,
{\it a priori}, the energy density of cold dark and baryonic matters no
longer scales as $1/a^3$ but as
\begin{equation}
\label{eq:cdmenergy}
\rho \sim \frac{1}{a^3} \sum _a n_a m^{_{\rm F}}_{\rm u,d}\left(
\frac{Q}{m_{_{\rm Pl}}}\right)\, ,
\end{equation}
where $n_a$ is the number of non-relativistic particles. It has been
shown that this type of interaction between dark matter and dark energy
can result in an effective dark energy equation of state less than
$-1$~\cite{DCK,MSU}. 

\par

Another consequence is the so-called chameleon
effect~\cite{KW,KW2,cham,BMcham}. The equation~(\ref{eq:cdmenergy})
implies that the effective potential for the quintessence field is
modified by matter and becomes $V_{\rm eff}(Q)= V_{_{\rm DE}}(Q) + f(Q)
\rho_{\rm mat}$. This potential is explicitly time-dependent and usually
possesses a minimum if $V(Q)$ has a runaway shape and $f(Q)$ is
increasing with $Q$. At the minimum, the time-dependent mass is given by
$m^2_{_{\rm Q}}=\left(V_{\rm eff}\right)_{,QQ}$. This implies two new
effects. Firstly, the mass of the field will depend on the environment
(hence the name chameleon) through $\rho_{\rm mat}$. If $\rho _{\rm
mat}$ is large then the mass can be such that $m_{_{\rm
Q}}>10^{-3}\mbox{eV}$, thus evading all the constraints on the fifth
force or on the weak equivalence principle. Secondly, there is the
so-called thin shell effect. If we consider a situation where the
gravitational experiments are performed on a spherical body of radius
$R_{\rm b}$ embedded in a surrounding medium, then the above effective
potential is not the same inside the body and outside because $\rho
_{\rm matter}$ is different. As a matter of fact, this modifies the
gravity tests. Indeed, one can calculate the profile of the quintessence
field inside and outside the body and show that the acceleration felt by
a test particle is given by
\begin{equation}
a=-\frac{Gm_{\rm b}}{r^2}\left[1+\frac{\alpha_{_{\rm Q}}
\left(Q_{\infty}-Q_{\rm b}\right)}{\Phi _{_{\rm N}}}\right]\, ,
\end{equation}
where $\Phi _{_{\rm N}}=Gm_{\rm b}/R_{\rm b}$ is the Newtonian potential
at the surface of the body. $Q_{\rm b}$ and $Q_{\infty}$ denote the
value of the field inside and outside the body respectively and the
quantity $\alpha _{_{\rm Q}}$ has been defined in
Eq.~(\ref{eq:defalpha}). Therefore, we see that the parameter which
controls the deviation from the Newtonian acceleration is
\begin{equation}
\label{eq:chamalpha}
\frac{\alpha _{_{\rm Q}}
\left(Q_{\infty}-Q_{\rm b}\right)}{\Phi _{_{\rm N}}}\, ,
\end{equation}
while, in absence of the thin shell mechanism, this would be $\alpha
_{_{\rm Q}}$ [notice that the limit $Q_{\infty }-Q_{\rm b}\rightarrow 0$
cannot be taken in Eq.~(\ref{eq:chamalpha}) because the presence of the
thin shell was assumed from the very beginning]. Hence, even if $\alpha
_{_{\rm Q}}$ is quite large, which would {\it a priori} rule out the
corresponding model, if the new factor $\left(Q_{\infty}-Q_{\rm
b}\right)/\Phi _{_{\rm N}}$ is small then the model can be compatible
with local tests of gravity. 

\section{Conclusions}
\label{sec:conclusions}

In this article, we have presented a short review on quintessence. The
main conclusion is that, from the high energy physics point of view, it
is difficult to build, at the same time, a model which is interesting
from the cosmological point of view (different from a pure cosmological
constant) and compatible with the local gravity tests.

\section*{Acknowledgments}

I would like to thank my collaborator P.~Brax and M.~Lemoine for careful
reading of the manuscript.

\end{document}